\documentclass[]{piparticle}

\usepackage[utf8]{inputenc}
\usepackage{graphicx}
\usepackage{bm}
\usepackage{epsfig}
\usepackage{amsmath}
\usepackage{amssymb}
\usepackage{amsfonts}
\usepackage{latexsym}
\usepackage{xcolor}
\usepackage{ulem}

\usepackage[inline]{trackchanges}
\addeditor{LP}
\addeditor{MM}
\addeditor{RD}

\newcommand{\Ds}{D_{\rm s}}
\newcommand{\Rs}{R_{\rm s}}
\newcommand{\Ro}{R_{\rm o}}
\newcommand{\muI}{\mu_o}
\newcommand{\Do}{D_{\rm o}}
\newcommand{\As}{A_{\rm s}}

\newcommand{\rhob}{\rho_{\rm b}}
\newcommand{\rhoo}{\rho_{\rm o}}
\newcommand{\Mp}{M_{\rm p}}
 
 \newcommand{\szzo}{\sigma_{zz}^o}
 \newcommand{\Qi}{Q_{\rm ini}}
 \newcommand{\Qf}{Q_{\rm end}}

\begin{document}

\runningauthor{L. A. Pugnaloni \itshape{et al.}}  

\title{Forced silo discharge: Simulation and theory}

\author{Luis. A. Pugnaloni,\cite{inst1}\thanks{E-mail: luis.pugnaloni@exactas.unlpam.edu.ar}  
        Marcos A. Madrid,\cite{inst2} 
        J. R. Darias\cite{inst3,inst4}}

\pipabstract{
We study, through discrete element simulations, the discharge of granular materials through a circular orifice on the base of a cylindrical silo forced by a surcharge. At the beginning of the discharge, for a high granular column, the flow rate $\Qi$ scales as in the Beverloo equation for free discharge. However, we find that the flow rate $\Qf$ attained at the end of the forced discharge scales as $\sqrt{\rhob P}\Do^3/\Ds$, with $\rhob$ the bulk density, $P$ the pressure applied by the overweight, $\Do$ the orifice diameter and $\Ds$ the silo diameter. We use the work--energy theorem to formulate an equation for the flow rate $\Qf$ that predicts the scalings only in part. We discuss the new challenges offered by the phenomenology of strongly forced granular flows.
}

\maketitle

\blfootnote{
\begin{theaffiliation}{99}
   \institution{inst1} Departamento de F\'isica, Facultad de Ciencias Exactas y Naturales, Universidad Nacional de La Pampa, CONICET, Uruguay 151, 6300 Santa Rosa (La Pampa), Argentina.
   \institution{inst2} Departamento de Ingenier\'ia Mec\'anica, Facultad Regional La Plata, Universidad Tecnol\'ogica Nacional, CONICET, Av. 60 Esq. 124, 1900 La Plata, Argentina.
   \institution{inst3} Laboratorio de Sistemas Complejos, Universidad Sim\'on Bol\'ivar, Apartado Postal 89000, Caracas 1080--A, Venezuela.
   \institution{inst4} Departamento de F\'isica, Universidad Metropolitana, Caracas, Venezuela.
\end{theaffiliation}
}

\section{Introduction}
\label{intro}

It is well known that granular materials in a silo discharge through an orifice, under the sole action of gravity, at a constant mass flow rate; unlike inviscid fluids \cite{hagen1852,BrownBook}. In such free discharge conditions, the mass flow rate $Q$ scales as $D_{\rm o}^{5/2}$, with $D_{\rm o}$ the diameter of the circular orifice \cite{hagen1852,beverloo}. In addition, the flow rate does not appear to be correlated with the local pressure around the discharge orifice \cite{aguirre2010,aguirre2011}. The Hagen--Beverloo equation for free discharge states
\begin{equation}
    Q = C \rhob \sqrt{g} (\Do - kd)^{5/2}.
\end{equation}
Here, $d$ is the mean particle diameter, $g$ the magnitude of acceleration of gravity, and $\rhob$ the bulk density. The parameters $C$ and $k$ are generally fitted to the experimental data. In general, $k\approx 1.5$ is suitable for spherical grains and $C\approx 0.58$ is a well established universal value with recent theoretical support \cite{darias2020}. 

It is particularly interesting that $Q$ does not depend on silo diameter $\Ds$ as long as $\Ds > 2\Do$. Besides, the material properties of the grains only come into the equation through $d$ and the density $\rho$ of the material the grains are made of. $Q$ is directly proportional to $\rho$ because $\rhob=\rho\phi$, with $\phi$ the packing fraction of the granular column. We note the contrast with liquids, whose mass flow rate depends on $\sqrt{\rho}$. The fact that $Q$ in a free discharge does not depend on other properties of the material, such as the Young modulus, the friction coefficient or the Poison ratio, has been explained in \cite{darias2020,madrid2018}. 

When the free surface of the granular column in the silo is pressed by an overweight (e.g., a heavy piston), the flow rate accelerates over the final stages of the discharge \cite{madrid2018,madrid2019}. The presence of the overweight leads to a pressure distribution in the silo that  {has recently been experimentally investigated} and contrasted with early models \cite{ferreyra2024}. Interestingly, forcing leads to flow rates that depend on material properties other than just $\rho$ \cite{madrid2018}. 

Apart from using a simple overweight to force the flow, researchers have tested a suspended-wall silo with overweights \cite{Peng2021} and the use of a layer of heavier grains on top of the granular column \cite{Horabik2022}. These two ways of forcing the flow do not show an acceleration of the flow during discharge but a constant flow rate (over the discharge time) that increases with the applied pressure. Interestingly, while in \cite{Peng2021} the flow rate grows linearly with the pressure at the base, in \cite{Horabik2022} it increases with the square root. There exists still a lack of experimental evidences, theoretical models and consensus regarding the phenomenology of forced silo discharges.

In this work, we performed simulations using the Discrete Element Method (DEM) to explore the mass flow rate $Q$ for a forced silo discharge as a function of $\Ds$, $\Do$, $\rho$ and pressure $P$ exerted by an overweight. We show that there exist two limiting cases: (i) at the beginning of the discharge, when the column of grains is very tall, the flow rate $\Qi$ is similar to that of Beverloo without any effect due to the overweight; (ii) at the end of the discharge, the flow rate $\Qf$ depends on the applied pressure and material density in a way similar to that of a liquid. Interestingly, the scaling with orifice diameter and silo diameter is unexpected for $\Qf$. By means of the work--energy theorem, we derive a differential equation for the mass flow rate $Q$ as a function of time during a forced discharge. To this end, we followed an approach similar to that used in \cite{darias2020} for unforced discharges and the stress model experimentally validated in \cite{ferreyra2024} for forced flows. We show that the theory predicts the existence of both limiting cases and part of the scalings for $\Qf$. We discuss some challenges faced in modeling strongly forced granular flows.  

\section{Simulations}

\begin{figure}
 \begin{center}
  \includegraphics[width=0.6\columnwidth]{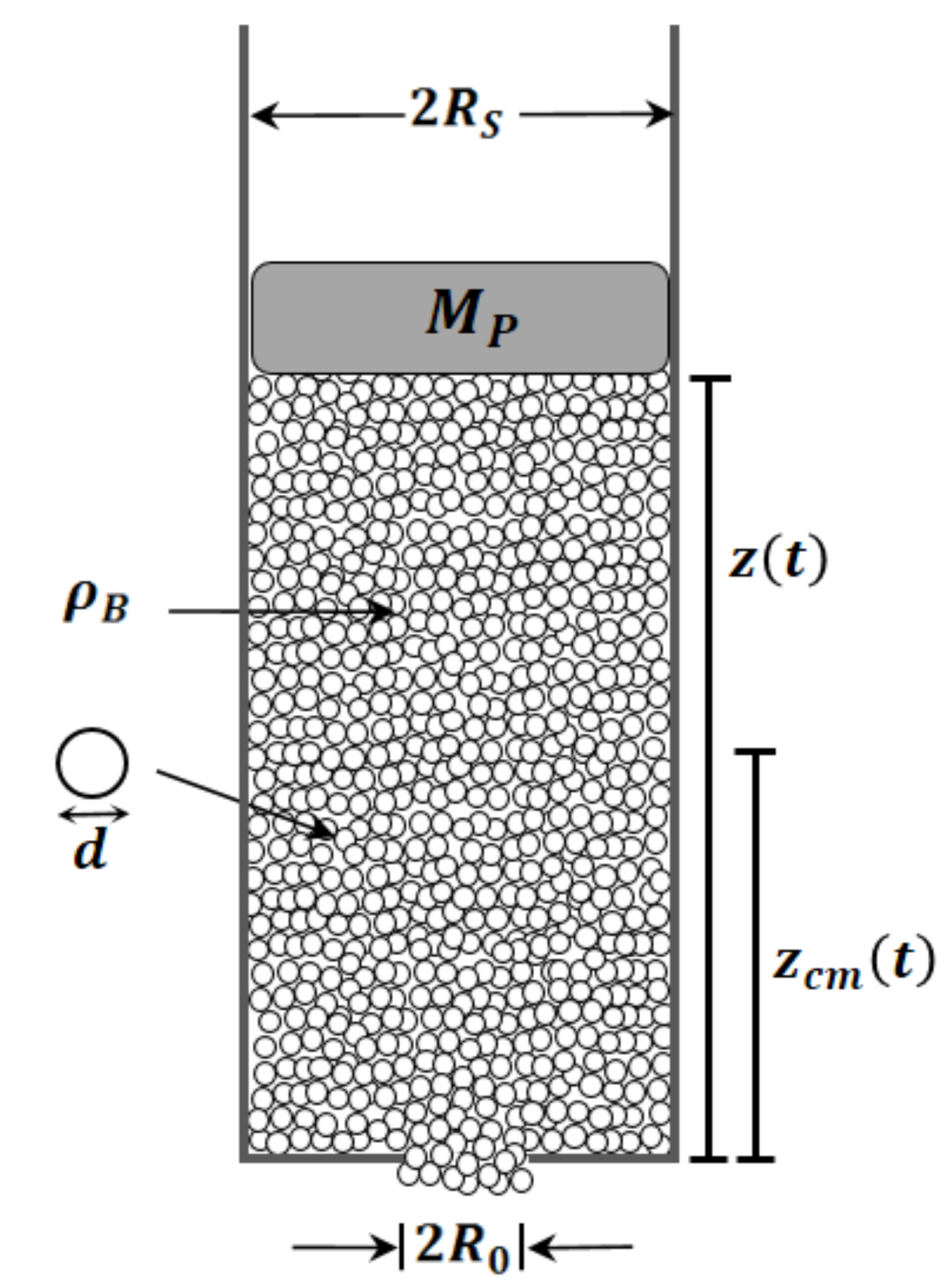}
 \end{center}
\caption{Sketch of the axial cross section of a cylindrical silo with granular material and overweight.}
\label{set-up}
\end{figure}

We consider a cylindrical silo (diameter $\Ds$) with a flat base as the one sketched in Fig. \ref{set-up}. The base has an orifice of diameter $\Do$ in its center. The silo is filled with an initial mass $M_{\rm ini}$ of a granular material (each grain has mass $m$ and diameter $d$) and a piston of mass $M_{\rm p}$ is placed as an overweight on top of the free surface. The piston is slightly narrower than the silo so that grains cannot fit into the gap between the piston and the silo walls.

We run DEM simulations of the forced silo discharge. We utilized LIGGGHTS \cite{liggghts} with a particle--particle Hertz interaction and Coulomb criterion using a Young modulus $Y=70$ MPa, Poisson ratio $\nu=0.25$, restitution coefficient $e=0.95$ and friction coefficient $\mu = 0.5$. Unless otherwise stated, the same interaction applies for the particle--walls contacts. In terms of dimensional units the particle diameter is set to $d=1.0$ mm. Particles (up to $3 \times 10^5$ grains, depending on the silo diameter) are poured into the silo with the orifice initially blocked by a plug. Then, a piston of desired weight $\Mp$ is placed on top of the granular column. The piston has no friction with the silo walls. We let the grains rest in the silo  waiting until the kinetic energy per particle falls below $10^{-10}$~J. Then, we remove the plug and allow the material to discharge through the orifice. The magnitude of the acceleration of gravity is set to $g=9.81$ m/s$^{2}$ and the integration time step is $\Delta t=5\times10^{-6}$ s.  Different silo diameters, grain material density, orifice diameter, and piston weight are tested. During the entire discharge, we record the particle positions, velocities, accelerations, and particle stress tensor.

\section{Results}
\label{sec:results}

\begin{figure}
    \centering
    \includegraphics[width=\linewidth]{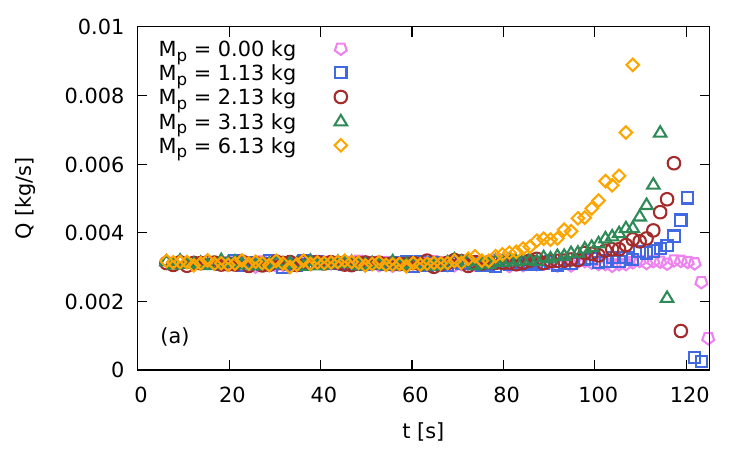}
    \includegraphics[width=1.05\linewidth]{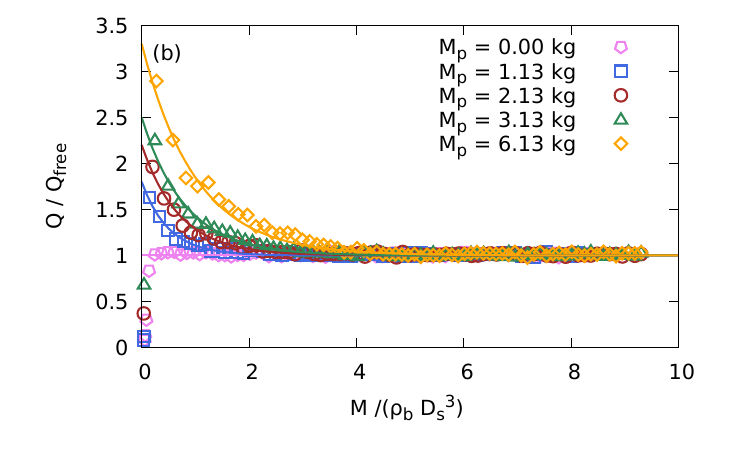}
    \caption{Mass flow rate $Q$ as a function of time (a) and as a function of the instantaneous mass inside the silo (b) for various overweight masses $\Mp$. In part (b) $Q$ is scaled by the mass flow rate $Q_{\rm free}$ of a free discharge through the same orifice. Simulations correspond to $\Ds=30d$, $\Do=6d$ and $\rho=2500$ kg/m$^3$. The solid lines correspond to fitting the function $Q(M) = (\Qf-\Qi) \exp[-M/M_{\rm decay}]+ \Qi $.}
    \label{fig:time}
\end{figure}

In Fig.~\ref{fig:time}, we present some sample discharge flow rate $Q$ as a function of time and as a function of the mass $M$ of grains inside the silo (a proxy for the column height). As in previous studies \cite{madrid2018}, we observe a constant flow rate at the beginning of the discharge followed by a rapid increase when the column height decreases. At the very end of the discharge, a sharp drop to zero flow is observed due to the fact that the silo runs out of particles. In Fig.~\ref{fig:time}(b), we have normalized $Q$ by the one obtained in a free discharge (i.e., for $\Mp=0$). It is clear that at the beginning of the discharge $Q=Q_{\rm free}$ and the forcing due to $\Mp$ does no matter. 

To characterize the two limiting flow rate values, we extract the initial flow rate $\Qi$ at very tall columns and the final flow rate $\Qf$ at zero column height (at the end of the acceleration process, just before the final drop to zero). To this end, we fit the simulations curves $Q$ vs. $M$ using the exponential function $Q(M) = (\Qf-\Qi) \exp[-M/M_{\rm decay}]+ \Qi$ and extract the two limits $\Qi=Q(M\rightarrow \infty)$ and $\Qf=Q(M\rightarrow 0)$ (see solid lines in Fig.~\ref{fig:time}(b)). We ran simulations for various $\Do$, $\Ds$, $\rhob$, and $\Mp$ and extracted $\Qi$ and $\Qf$ in each case. 

\begin{figure}
    \centering
    \includegraphics[width=\linewidth]{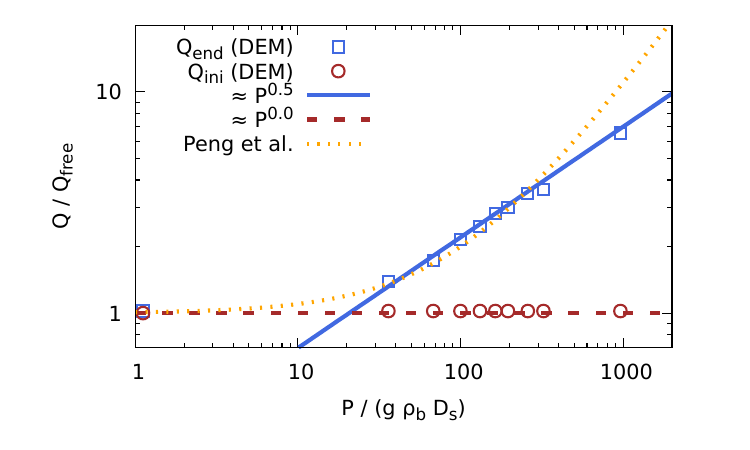}
    \caption{Mass flow rate for a tall column $\Qi$ and at the end of the discharge $\Qf$ scaled by that of a free discharge $Q_{\rm free}$ as a function of the pressure $P$ applied by the overweight. Simulations correspond to $\Do=6d$, $\Ds=30d$ and $\rhob=2500$ kg/m$^3$. The yellow dotted line corresponds to Peng et al. model \cite{Peng2021}. Thick lines correspond to different power laws.}
    \label{fig:presion}
\end{figure}

First, we consider the effect of the applied pressure $P=4\Mp/\pi\Ds^2$ as shown in Fig.~\ref{fig:presion} for given $\Do$, $\Ds$ and $\rhob$. As we mentioned above, the initial flow rate $\Qi$ is independent of $P$. However, pressure induces an increase in the final flow rate $\Qf$ as seen by others \cite{madrid2019,Peng2021,Horabik2022}. In Fig.~\ref{fig:presion} we have used log--log scale to highlight the power law at high pressures, which is compatible with $\sqrt{P}$. This is in accordance with the discussion presented by Horabik et al.~\cite{Horabik2022} but contrasts with the linear dependency suggested by Peng et al.~\cite{Peng2021}. A remarkable feature is that this square root scaling with $P$ is akin to the one observed in viscous fluids, while a free discharge presents a flow rate independent of pressure \cite{aguirre2010,NeddermanBook}. It is worth mentioning that at low forcing pressures, $\Qf$ tends to $Q_{\rm free}$, as observed in experiments \cite{madrid2019,Peng2021}, and the square root scaling ceases. Peng et al. proposed a base flow rate at zero pressure which corresponds to the free discharge corrected by a linear term $\alpha P$. This empirical equation works well at low pressures (see yellow dotted line in Fig.~\ref{fig:presion}) but fails at large $P$. We will show in Section \ref{sec:balance} that the Work-Energy theorem can predict the observed square root dependency for high $P$.

\begin{figure}
    \centering
    \includegraphics[width=\linewidth]{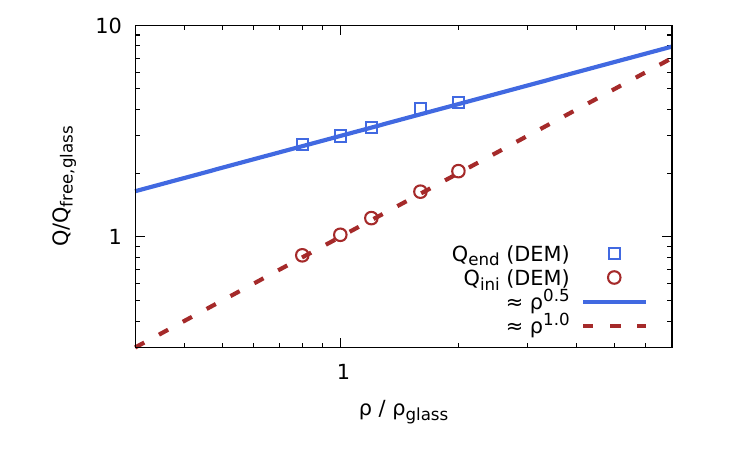}
    \caption{$\Qf$ and $\Qi$ scaled by $Q_{\rm free}$ of glass as a function of the material density of the particles $\rho$ scaled by the density of glass ($\rho_{\rm glass}=2500$ kg/m$^3$). Simulations correspond to $\Do=6d$, $\Ds=30d$ and $P=85.0$ kPa. Lines correspond to different power laws.}
    \label{fig:densidad}
\end{figure}

In view of the previous result, which indicates that $\Qf$ resembles a viscous flow, we consider the effect of density. We recall that, in a free discharge, the mass flow rate is proportional to the density of the granular material \cite{beverloo}, whereas in a viscous flow it is proportional to $\sqrt{\rho}$. Figure~\ref{fig:densidad} shows $\Qf$ and $\Qi$ as a function of $\rho$ on a logarithmic scale. As we can see, the scaling is also compatible with a viscous flow with $\Qf \propto \sqrt{\rho}$ while $\Qi \propto \rho$. 

\begin{figure}
    \centering
    \includegraphics[width=\linewidth]{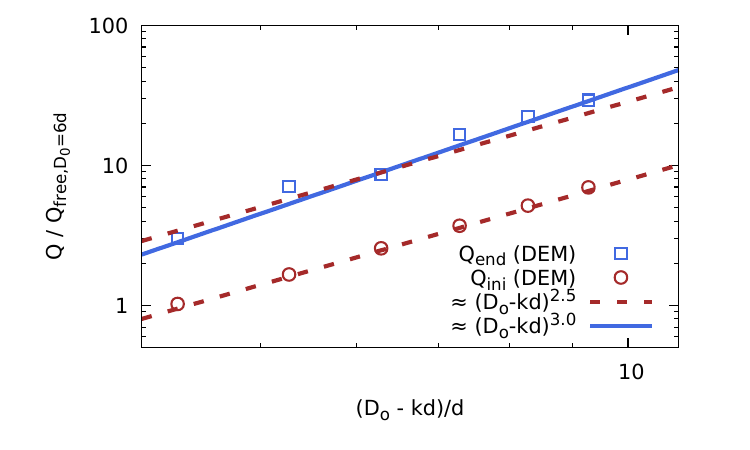}
    \caption{$\Qf$ and $\Qi$ scaled by $Q_{\rm free}$ at $\Do=6d$ as a function of the empty annulus corrected orifice diameter $\Do-kd$ scaled by $d$. Simulations correspond to $\rho=2500$ kg/m$^3$, $\Ds=30d$ and $P=85.0$ kPa. Lines correspond to different power laws.}
    \label{fig:orificio}
\end{figure}

Having observed a viscous-like behavior for $\Qf$ with $P$ and $\rho$, we explore the effect of the orifice diameter that we would expect to present a quadratic form as in a fluid. Figure~\ref{fig:orificio} shows $\Qi$ and $\Qf$ as a function of $\Do-kd$. We have introduced the empty annulus correction $kd$ as done by Beverloo \cite{beverloo}. Surprisingly, the scaling is compatible with $\Qf \propto (\Do-kd)^3$. This is neither the expected quadratic dependency observed in fluids nor the $5/2$ power observed in a free granular discharge. It is hard to tell apart power laws with exponents $3$ and $5/2$. However, our data seems to be better fitted by $\Qf \propto (\Do-kd)^3$. As a reference, we also include in Fig.~\ref{fig:orificio} the flow rate for tall columns $\Qi$. In this case, the forcing pressure does not affect the flow, which remains the same as for a free discharge, and the observed $5/2$ scaling is compatible with the Beverloo scaling. In the next section, we will propose a plausible explanation for the origin of this unexpected scaling. 

\begin{figure}
    \centering
    \includegraphics[width=\linewidth]{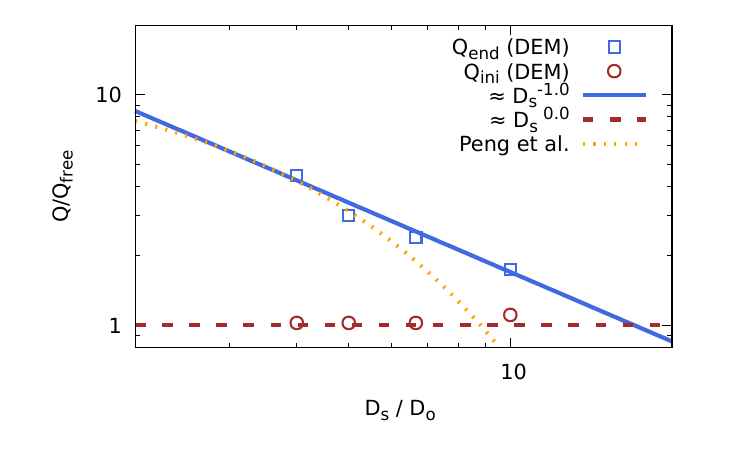}
    \caption{$\Qf$ and $\Qi$ scaled by $Q_{\rm free}$ as a function of the silo diameter $\Ds$ scaled by $\Do$. Simulations correspond to $\rho=2500$ kg/m$^3$, $\Do=6d$ and $P=85.0$ kPa. The yellow dotted line corresponds to Peng et al. \cite{Peng2021}. The thick lines correspond to different power laws.}
    \label{fig:silo}
\end{figure}

Finally, we consider the effect of the silo diameter $\Ds$. Although in a free discharge $\Ds$ is not relevant except for silos in which $\Do$ approaches $\Ds$ \cite{mankoc2007}, we expect some effect in forced silos based on previous studies \cite{Peng2021}. In Fig.~\ref{fig:silo}, we present $\Qi$ and $\Qf$ as a function of $\Ds$. As we expected, $\Qi$ is independent of $\Ds$. However, $\Qf$ falls as $\Ds^{-1}$. Peng et al. found an exponential decay, for a suspended-wall forced silo, rather than a power law \cite{Peng2021}. We will show that the work--energy theorem predicts a power-law dependency on $\Ds$, but with a different exponent. However, we note that this power law decay has to break down at very large $\Ds$ since $\Qf$ must necessarily converge to a constant limiting value $Q_{\rm free}$ set by a free discharge.

\begin{figure}[b]
    \centering
    \includegraphics[width=\linewidth]{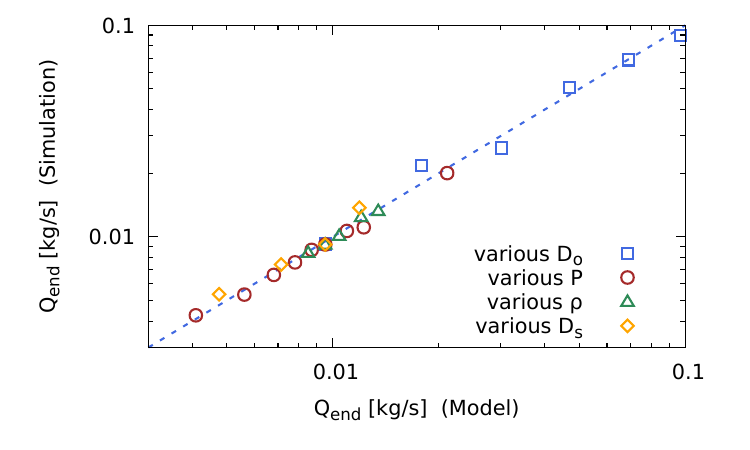}
    \caption{$\Qf$ from simulation versus the model scaling of Eq.~(\ref{eq:scaling}) with $\chi=0.25$ for a range of material densities, orifice diameter, silo diameter and applied overweight pressure.}
    \label{fig:scaling}
\end{figure}

To summarize, in a forced discharge, while $\Qi$ conserves all the scalings known for a free discharge, the behavior of $\Qf$ seems to be compatible with

\begin{equation}\label{eq:scaling}
    \Qf = \chi \sqrt{\rhob P} \frac{(\Do-kd)^3}{\Ds}.
\end{equation}
This scaling works for $\Ds<10\Do$ and $P>10 g \rhob \Ds$. According to our simulation results $\chi \approx 0.25$. Figure~\ref{fig:scaling} collects the $\Qf$ values for the range of control parameters investigated and compares them against Eq.~(\ref{eq:scaling}).

\section{Work-energy theorem}
\label{sec:balance}

Consider the silo sketched in Fig. \ref{set-up}. Let $D_{\rm s}$ be its diameter, $\Rs=D_{\rm s}/2$ its radius, and $\As=\pi \Rs^2$ its cross section. The base has an orifice of diameter $D_{\rm o}$ (radius $R_{\rm o}=D_{\rm o}/2$ and cross section $A_{\rm o}=\pi R_{\rm o}^2$) in its center. The silo is filled with an initial mass $M_{\rm ini}$ of a granular material (each grain has mass $m$) and a piston of mass $M_{\rm p}$ is placed as an overweight on top of the free surface. The grains cannot fit into the gap between the piston and the silo walls. As an approximation, we assume that the bulk density $\rhob$ of the granular material is homogeneous. 

During the discharge, the mass $M(t)$ inside the silo at time $t$ can be written as
\begin{equation}{\label{equiv}}
 M(t)=\rhob \As z(t)=2\rhob \As z_{\rm cm}(t),
\end{equation}
where $z(t)$ is the height of the column of grains, and $z_{\rm cm}(t)=z(t)/2$ is the vertical position of the center of mass of the granular column at time $t$. Therefore, the mass flow rate $Q(t)$ is
\begin{equation}
 Q(t)=-\dot{M}(t)=2 \rhob \As v_{\rm cm}(t), \label{q-vcm}
\end{equation}
where $v_{\rm cm}(t)=|\dot{z}_{\rm cm}(t)|$ is the speed of the center of mass of the granular column. We note that since $M(t)$ decreases with time during the discharge, $\dot{M}$ and $\dot{z}_{\rm cm}$ are negative.

Let us consider the system composed of the grains inside the silo at any given time. The work--energy theorem states that the change in kinetic energy $\dot{K}_{\rm in}$ of this system is 
\begin{equation}
\dot{K}_{\rm in} = W_{\rm g} + W_{\rm p} - W_{\rm out} - W_{\rm el} - W_{\rm d}, \label{energy-balance}
\end{equation}
where $W_{\rm g}$ is the power injected by the force of gravity acting on the grains, $W_{\rm p}$ is the power injected by the weight of the piston, $W_{\rm out}$ is the power loss due to the grains that leave the silo through the orifice at a velocity $v_{\rm out}$, $W_{\rm el}$ is the ``elastic power'', i.e.,  the rate of change of the elastic energy of the grain--grain contacts, and $W_{\rm d}$ is the dissipated power due to the non--conservative interactions (friction and inelastic collisions between the grains and between the grains and the walls). In the following subsections we will consider in detail each of the terms of Eq.~(\ref{energy-balance}).

\subsection{Internal kinetic energy ($\dot{K}_{\rm in}$)}

\begin{figure}
 \begin{center}
\includegraphics[width=1\columnwidth]{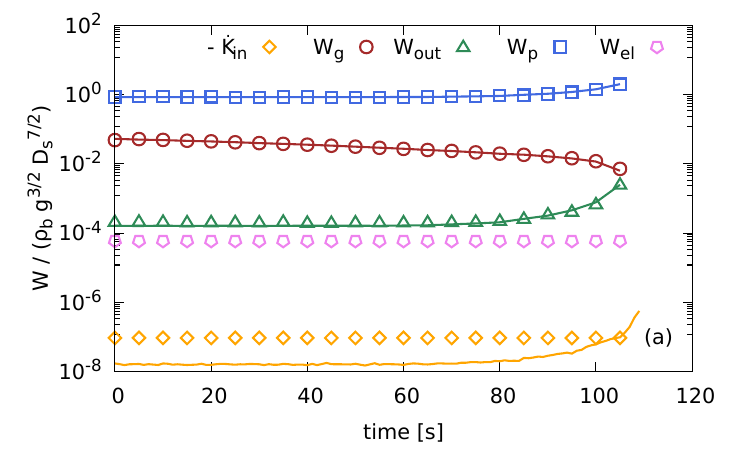}
\includegraphics[width=1\columnwidth]{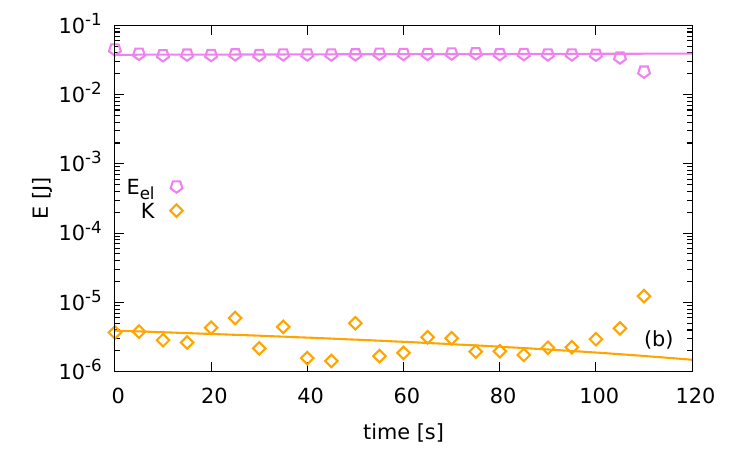}
 \end{center}
\caption{(a) Power contributions during the discharge of a forced silo with $\Mp=6.0$ kg as a function of time (DEM simulations). (b) Smoothed kinetic energy $K_{\rm in}(t)$ and elastic energy $E_{\rm el}(t)$ with linear fits to estimate $\dot{K}_{\rm in}$ and $W_{\rm el}$ from the slope, shown as  constant values in part (a). The solid lines in part (a) correspond to the theoretical estimates for: $W_{\rm p}(t)$ [Eq. (\ref{wp}), blue]; $W_{\rm g}(t)$ [Eq. (\ref{wg}), maroon]; $W_{\rm out}(t)$ [Eq. (\ref{wout}), green] and $\dot{K}_{\rm in}(t)$ [second term of Eq. (\ref{k-in}), yellow].}
\label{energies}
\end{figure}

Figure \ref{energies}(a) displays different power contributions extracted from a DEM simulation during a forced silo discharge. The fluctuations in kinetic energy $K_{\rm in}$ and elastic energy $E_{\rm el}$ are large relative to the mean, which makes time derivatives to calculate the power not very reliable. Therefore, to estimate $\dot{K}_{\rm in}$ and $ W_{\rm el}$ we plot $K_{in}(t)$ and $E_{\rm el}(t)$ as functions of time in Figure \ref{energies}(b) and use the slope of a fitting straight line as a mean time derivative. As observed in Fig.~\ref{energies}(b), neither $K_{\rm in}$ nor $E_{\rm el}$ are linear in time at the end of the discharge, but we fit only the first half of the discharge time to have a representative slope. Simulations using other silo diameters, overweight and material density display the same trends as observed for free discharge in \cite{darias2020}. As we can see in Fig. \ref{energies}, the contribution of $\dot{K}_{\rm in}$ is orders of magnitude smaller than any other contribution. Based on this, we will neglect $\dot{K}_{\rm in}$ in Eq. (\ref{energy-balance}). To provide a partial explanation for this observation, consider the kinetic energy of the grains inside the silo
\begin{equation}
 K_{\rm in}(t)=\frac{1}{2}\sum_{i=1} m \bm{v}_i^2(t).
\end{equation}
The sum runs over all particles inside the silo at time $t$ and $\bm{v}_i(t)$ is the velocity of particle $i$. This can be expressed in terms of the center of mass velocity $v_{\rm cm}$  and the ``granular temperature'' as $K_{\rm in}(t)= K_{\rm in}^{\rm cm}(t) + K_{\rm in}^{\rm temp}(t)$, being
\begin{equation} \label{kinetic-cm2}
 K_{\rm in}^{\rm cm}(t)=\frac{1}{2}M(t) v_{\rm cm}^2(t),
\end{equation}
and
\begin{equation} \label{kinetic-cm3}
 K_{\rm in}^{\rm temp}(t)= \frac{1}{2}\sum_{i=1}^{N(t)}m_i [\bm{v}_i(t)-\bm{v}_{\rm cm}(t)]^2.
\end{equation}

We will disregard the term ``temperature'' and focus on the contribution of the center of mass. We assume here that the $x$ and $y$ components of $\bm{v}_{\rm cm}$ are null and, therefore, $|\bm{v}_{\rm cm}|=v_{\rm cm}=-\dot{z}_{\rm cm}$.
\begin{align}
\dot{K}_{\rm in}(t)&\approx \dot{K}_{\rm in}^{\rm cm}(t) \notag\\&=M(t) v_{\rm cm}(t) \dot{v}_{\rm cm}(t) + \frac{1}{2}v_{\rm cm}^2(t) \dot{M}(t). 
\end{align}
which can be written, using Eqs. (\ref{equiv}) and (\ref{q-vcm}), as
\begin{equation}
 \dot{K}_{\rm in}(t)\approx \frac{1}{4\rho^2_{\rm b} A^2_{\rm s}}
 \left[M(t)\dot{M}(t)\ddot{M}(t)+\frac{1}{2}\dot{M}^3(t)\right] \label{k-in}.
\end{equation}
As we can see, $\dot{K}_{\rm in}$ decays as $\As^{-2}$. Therefore, $\dot{K}_{\rm in}$ will be small for wide silos. We will see below that other terms in the energy balance fall more smoothly with $\As$. Hence, $\dot{K}_{\rm in}$ can be neglected as a first approximation. In practice, we can neglect only the first term in the brackets of Eq. (\ref{k-in}) and keep the second term. This contribution is still small for large $\Ds$ and it can be proved that it corresponds to the known correction factor of the flow rate of an orifice-plate configuration for fluids given by $1/\sqrt{1-(\Ds/\Do)^4}$  \cite{Reader2015}. 

In Fig. \ref{energies}(a), we test the prediction of the second term in Eq. (\ref{k-in}). To plot this expression, we use in Eq. (\ref{k-in}) the instant value of $\dot{M}(t)$ measured in the simulation. As we can see, the prediction is far from perfect, but gives the correct order of magnitude and allows us to justify neglecting $\dot{K}_{\rm in}$ in the energy balance.

\subsection{Gravitational energy ($W_{\rm g}$)}

The gravitational potential energy of the particles inside the silo is
\begin{equation}
 U_{\rm g}(t)=M(t)gz_{\rm cm}(t)=\frac{g M^2(t)}{2\rhob \As}.
\end{equation}
Where we have used Eq. (\ref{equiv}). Therefore, the power injected by the action of gravity is
\begin{equation}
 W_{\rm g}(t)=-\dot{U}_{\rm g}(t)=-\frac{g}{\rhob \As} M(t) \dot{M}(t). \label{wg}
\end{equation}
As we can see, $W_{\rm g}$ scales with $\As^{-1}$ in contrast to the faster decay displayed by $\dot{K}_{\rm in}$. We recall here that $\dot{M}(t)$ is negative; therefore $W_{\rm g}(t)$ is positive. In Fig. \ref{energies}(a), we show that Eq. (\ref{wg}) is consistent by using the instant values of $M(t)$ and $\dot{M}(t)$ measured in the simulation.

\subsection{Overweight energy ($W_{\rm p}$)}

The force applied by the piston on the granular column is
\begin{equation}
 F_{\rm p} = M_{\rm p} (g - a_{\rm p}(t)) = M_{\rm p} (g - \ddot{z}(t)). 
\end{equation}
Here, $a_{\rm p}$ is the acceleration experienced by the piston during discharge. The power injected can be calculated as
\begin{equation}
 W_{\rm p}(t)= - F_{\rm p} \dot{z}(t) = M_{\rm p} (g - \ddot{z}(t))\dot{z}(t).
\end{equation}

Using Eq. (\ref{equiv}), the power injected by the piston is
\begin{equation}\label{eq:wp1}
 W_{\rm p}(t)= -\frac{g M_{\rm p}}{\rhob \As}  \dot{M}(t)+ \frac{M_{\rm p}}{\rhob^2 \As^2} \dot{M}(t)\ddot{M}(t).
\end{equation}

We have tested in simulations that the second term in Eq. (\ref{eq:wp1}) is small enough and can be neglected as a first-order approximation. Hence,
  \begin{equation}  \label{wp}
 W_{\rm p}(t)\approx -\frac{g M_{\rm p}}{\rhob \As}  \dot{M}(t).
\end{equation}
As shown in Fig. \ref{energies}(a), the expression in Eq. (\ref{wp}) is an excellent approximation.

\subsection{Discharge energy ($W_{\rm out}$)}

While the system discharges, the particles that leave the system take away some energy since their own kinetic energy is no longer part of the internal energy of the silo. If during a time interval $dt$ the system discharges a mass $dM= -\dot{M}(t)dt$ at velocity $v_{\rm out}$, then the kinetic energy removed per unit time is
\begin{equation}
\dot{K}_{\rm out}(t) = -\frac{1}{2} v_{\rm out}^2 \dot{M}(t).
\end{equation}

By definition, $-\dot{M}(t)= \rho_{\rm o} A_{\rm o} v_{\rm out}$. Therefore,
\begin{equation}
 W_{\rm out}=\dot{K}_{\rm out}(t) =-\frac{\dot{M}^3(t)}{2\rho^2_{\rm o} A^2_{\rm o}}. \label{wout}
\end{equation}

The discharge power does not depend on the silo cross section, but on the orifice cross section $A_{\rm o}$. Then again, $W_{\rm out}$ is positive since $\dot{M}(t)$ is negative.  In Fig. \ref{energies}(a), we show that Eq. (\ref{wout}) is consistent by using the instant value of $\dot{M}(t)$ measured in the simulation.

\subsection{Elastic energy ($W_{\rm el}$)}

If the grains are stiff, the variation in the elastic energy at the contacts is expected to be small. $W_{\rm el}$ corresponds to the rate of change of the conservative component of the contact forces $\dot{E}_{\rm el}$. In Fig. \ref{energies}(a), we show $W_{\rm el}$ during a forced silo discharge in a DEM simulation. As we can see, this term is small in comparison with the other energy contributions. Hence, we will neglect $W_{\rm el}$ as a first approximation. Of course, this may be inadequate for very soft grains. For our purposes, we simply disregard this contribution in the analysis. However, this may eventually be included by using the stress based expression for the elastic energy density in the limit of small deformations \cite{landau}  
\begin{equation}
 E_{\rm el} = A_\textbf{s} z_\text{cm} \bm{\sigma} \bm{\epsilon}, \label{elastic}
\end{equation}
being $\bm{\sigma}$ the stress tensor and $\bm{\epsilon}$ the strain tensor.

\subsection{Dissipated energy}

In previous works \cite{madrid2018,darias2020}, we have shown that the power dissipated during a silo discharge can be calculated by assuming that the flow is consistent with a quasistatic shear flow in the framework of the $\mu(I)$-rheology model \cite{dacruz2005}. Here, we present a derivation of $W_{\rm d}$ that considers in detail the effect of the convergent flow close to the orifice and the effect of the forcing piston. 

According to the $\mu(I)$-rheology model, in a plane shear configuration where a granular sample is sandwiched between two plates, the average tangential stress $\sigma_{xz}$ required to move the plates at constant relative velocity $v$ can be written as $\sigma_{xz}= \mu(I)\sigma_{zz}$. Here, $\sigma_{zz}$ is the confining stress pressing the two plates that constraint the particles and $\mu(I)$ is an ``effective friction coefficient'' that accounts for all the complex dissipative interactions in the granular sample between the plates. This effective friction depends on the inertial number $I=v d \sqrt{\rho}/(L \sqrt{\sigma_{zz}})$, with $\rho$ the density of the material of the grains, $d$ is the mean size of the grains and $L$ the distance between the plates. The inertial number suffices to characterize the flow as long as the grains are stiff and the plate-to-plate distance is much larger than the grain size. 

The power $W_{\rm d}$ dissipated during the motion of the plates in plane shear is simply given by $W_{\rm d}=\sigma_{xz} A v$, where $A$ is the total area of a plate. Hence, the dissipated power can be written as $W_{\rm d}=\mu(I)\sigma_{zz} A v$. In this expression, the effect of the properties of the granular material on dissipation comes only through $\mu(I)$ since $\sigma_{zz}$, $A$ and $v$ are set externally in the experiment. 

{For spherical grains, it has been shown that $\mu(I)$ follows a universal curve as long as the particle--particle friction coefficient $\mu$ is high enough (roughly $\mu > 0.4$) \cite{dacruz2005}. However, this curve depends on geometrical factors such as dimensionality \cite{azema2014}. Interestingly, in the quasistatic limit (i.e., $I\ll1$, with $I<0.01$ being a typical criterion used for practical purposes), the value of $\muI \equiv \mu(I=0)$ converges to a value that is fairly insensitive to material properties such as $\mu$ (provided that $\mu>0.4$), restitution coefficient and Young's modulus. Therefore, the dissipated power becomes fairly independent from the details of the particle--particle interactions for spherical grains.}

\begin{figure}
    \centering
    \includegraphics[width=0.4\columnwidth]{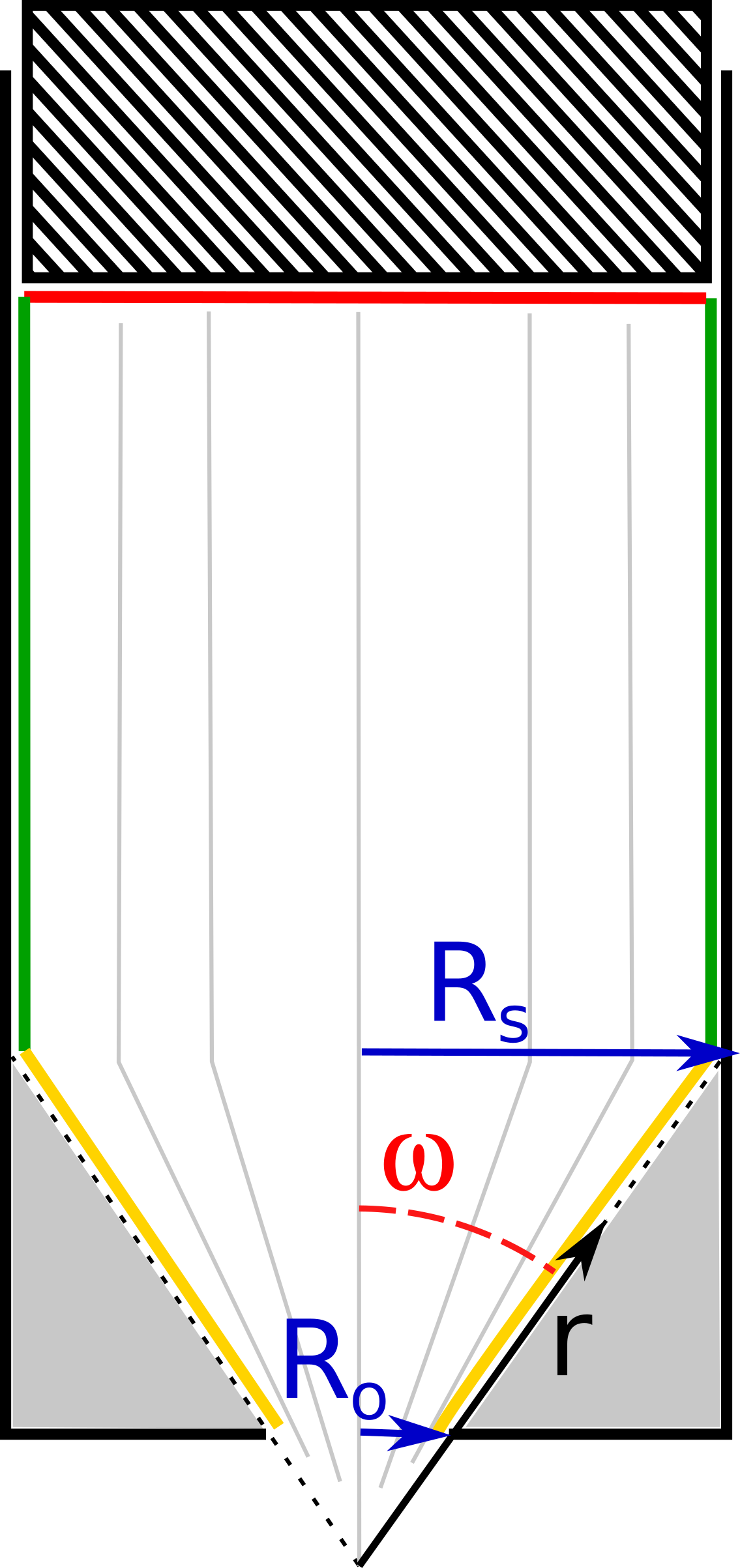}
    \caption{Streamlines during a forced discharge with a stagnant zone and convergent flow. The dissipation due to the shear on the three surfaces highlighted in red, green and yellow are calculated as separated contributions.}
    \label{fig:lineas1}
\end{figure}

We postulate that the flow inside a discharging silo can be modeled in a similar way to the simple plane shear geometry. {Let us consider the concentric shear due to the velocity gradient in the radial direction between the side vertical walls, highlighted in green in Fig.~\ref{fig:lineas1}, and the vertical axis of the silo.} In the cylindrical silo geometry, the confining pressure can be taken as $\sigma_{rr}$, {the characteristic driving velocity along the silo axis} as the velocity of the free surface of the granular column $v=2v_{\rm cm}$ and the ``plate area'' as the area of contact $\pi \Ds 2 z_{\rm cm}(t)$ between the grains and the lateral silo walls. Therefore, 
\begin{equation}
 W_{\rm d}^{\rm side\ wall}(t)= \muI  \langle\sigma_{rr}\rangle (t) [\pi \Ds 2 z_{\rm cm}(t)]  2 v_{\rm cm}(t)  . \label{eq:wd1}
\end{equation}
Here, $\langle\sigma_{rr}\rangle (t)$ is the mean normal wall pressure along the column of grains.

The region close to the base of the silo presents a convergent flow whose dissipation is not taken into account in Eq. (\ref{eq:wd1}). A stagnant cone is apparent with a slope that forms an angle $\omega$ with the vertical (see Fig.~\ref{fig:lineas1}). We can calculate the power dissipated due to the shear induced by the conical surface highlighted in yellow in Fig.~\ref{fig:lineas1} as
\begin{eqnarray}\label{eq:wcint}
    W^{\rm cone}_d(t) = \int_{\Ro/\sin\omega}^{\Rs/\sin \omega}{v_r(r)\sigma_{r\theta}(r) 2\pi r \sin\omega\, dr}. 
\end{eqnarray}

Here, we have used a spherical coordinate system with $r$ denoting the distance from the vertex of the cone (which lies below the orifice, outside the silo, see Fig.~\ref{fig:lineas1}). The velocity in the $r$-direction is denoted by $v_r(r)$ and the shear stress is denoted by $\sigma_{r\theta}(r)$. Due to the conical flow, mass conservation implies that $v_r(r)$ grows quadratically with decreasing $r$ (see \cite{NeddermanBook} \S 10.3). Hence,
\begin{eqnarray}\label{eq:vrr}
    v_r(r) = \frac{-\dot{M}(t)}{\Omega \rhob r^2}.
\end{eqnarray}

Where $\Omega=2\pi(1-\cos\omega)$ is the solid angle subtended by the cone. 

The shear stress $\sigma_{r\theta}(r)$ should decrease from a finite value $\sigma_o$, at the upper part of the cone, next to the wall, down to the orifice. As proposed by Neddermann (\cite{NeddermanBook} \S 10.3), we assume a linear decrease of the form
\begin{eqnarray}\label{eq:srt}
    \sigma_{r\theta}(r) = \sigma_o \frac{r \sin\omega}{\Rs}.
\end{eqnarray}

Plugging (\ref{eq:vrr}) and (\ref{eq:srt}) into (\ref{eq:wcint}) and integrating, we obtain
\begin{eqnarray}
    W^{\rm cone}_d(t) = - \sigma_o \frac{\pi\sin\omega}{1-\cos\omega} \Rs (\Rs-\Ro) \frac{\dot{M}(t)}{\rhob\As}.
\end{eqnarray}
It can be shown that $\sigma_o$ is proportional to the vertical stress $\szzo$ at the bottom of the silo and proportional to $\muI$ (see Appendix A). Therefore, we can collect proportionality constant for $\sigma_o$ and the factor $\pi\sin\omega/(1-\cos\omega)$ into a single constant $\alpha$ to write
\begin{eqnarray}\label{eq:wd2}
    W^{\rm cone}_d(t) = -\alpha \muI \szzo \Rs (\Rs-\Ro) \frac{\dot{M}(t)}{\rhob\As}. 
\end{eqnarray}
with $\alpha$ a constant that depends on the angle of the stagnant cone, the internal angle of friction, and the wall angle of friction (see Appendix A).

In addition to the dissipation due to the lateral walls and the lower conical flow, there is a contribution due to the contact between the forcing piston and the granular material (see the surface highlighted in red in Fig.~\ref{fig:lineas1}). In this case, the confining pressure due to the piston is $g\Mp/\As$,
with $\As$ the ``plate area''. The grains under the piston move tangential to it with a relative velocity that can be taken as $\gamma 2 v_{\rm cm}(t)$, with $\gamma$ a constant of order one. Therefore,
\begin{eqnarray}\label{eq:wd3}
    W^{\rm piston}_{\rm d}(t) = - \muI \gamma g \Mp \frac{\dot{M}(t)}{\rhob\As}. 
\end{eqnarray}
Summing Eqs. (\ref{eq:wd1}), (\ref{eq:wd2}) and (\ref{eq:wd3}) and using Eq. (\ref{q-vcm}), we obtain
\begin{align}
W_{\rm d}(t) = &W^{\rm side\ walls}_{\rm d} +W^{\rm cone}_{\rm d} +W^{\rm piston}_{\rm d} \notag \\ 
= - &\muI \langle \sigma_{rr}\rangle (t) 2 \pi \Rs \frac{M(t)}{\rhob\As} \frac{\dot{M}(t)}{\rhob \As} \notag \\
- &\muI \alpha \sigma^0_{zz}(t) 2 \pi \Rs(\Rs-\Ro) \frac{\dot{M}(t)}{\rhob \As} \notag\\
 - &\muI \gamma g M_{\rm p} \frac{\dot{M}(t)}{\rhob \As´}.
\label{wd}
\end{align}

\subsection{Differential equation for $\dot{M}(t)$}\label{sec-diff}

Collecting all terms for the energy balance from the previous section, i.e., plugging (\ref{wg}), (\ref{wp}), (\ref{wout}) and (\ref{wd}) into Eq. (\ref{energy-balance}), one can obtain a differential equation for the mass flow rate $-\dot{M}(t)$. As we mentioned before, some of these terms are in fact negligible. In particular, we take $\dot{K}_{\rm in}=0$ and $\dot{E}_{\rm el}=0$. Finally, Eq. (\ref{energy-balance}) can be written as
 \begin{align} \label{pre-diff}
 0 &=  -\frac{g}{\rhob \As} M(t) \dot{M}(t) 
 -\frac{g}{\rhob \As} M_{\rm p} \dot{M}(t)
       + \frac{\dot{M}^3(t)}{2\rho^2_{\rm o} A^2_{\rm o}} \notag\\
   & + \left[ \langle\sigma_{rr}\rangle (t) \frac{M(t)}{\rhob \As}+ \alpha \szzo(t)  (\Rs-\Ro) + \gamma \frac{gM_{\rm p}}{2 \pi \Rs} \right]  \notag\\
   & \times \muI 2 \pi \Rs \frac{\dot{M}(t)}{\rhob \As}. 
\end{align}
Solving for $-\dot{M}\equiv Q(t)$ \cite{foot}
\begin{eqnarray}  \label{diff}
-\dot{M}(t) = &\frac{\pi\sqrt{2}}{4}& \rho_{\rm o} \sqrt{g} D_{\rm o}^2  \left[ \frac{M(t)+M_{\rm p}}{\rhob \As} \right. \\
 &-& \frac{\muI 2 \pi \Rs}{g \rhob \As} \left(\langle\sigma_{rr}\rangle (t) \frac{M(t)}{\rhob \As}\right. \notag\\
  &+& \left. \left. \alpha \szzo(t)  (\Rs-\Ro)+ \frac{\gamma gM_{\rm p}}{2 \pi \Rs}\right) \right] ^{1/2}.\notag
\end{eqnarray}

Equation (\ref{diff}) is a first-order differential equation for $M(t)$ that can be closed with an initial condition such as $M(t=0)= M_0$. To solve this equation, it is necessary to know $\langle\sigma_{rr}\rangle (t)$, $\sigma^0_{zz}(t)$, $\muI$, $\alpha$, and $\gamma$. In the next section, we review Walters' model for the pressure in discharging silos to close Eq. (\ref{diff}).

It is worth mentioning at this point that Eq. (\ref{pre-diff}) reduces to the equation for an inviscid fluid if the last term that accounts for the dissipated power is neglected and $M_{\rm p}=0$. In this case, we obtain
\begin{eqnarray}  \label{bernoulli}
-\frac{\dot{M}(t)^2}{\rho^2_{\rm o} A^2_{\rm o}} &=& 2 g \frac{M(t)}{\rhob \As},\notag\\
v_{\rm out}&=& \sqrt{2 g h(t)},
\end{eqnarray}
where $h(t)=M(t)/(\rhob \As)$ and $v_{\rm out}=\dot{M}(t)/(\rho^2_{\rm o} A^2_{\rm o})$.

\subsection{Pressure from Walters model} \label{sec:walters}

Walker \cite{walker1966} and Walters \cite{walters1973} have developed models for the pressure during silo discharge (including an overweight) following a similar approach to that of Janssen \cite{janssen1895}. These developments went beyond in the sense that the Janssen's redirection constant was put in terms of the internal and wall effective friction. Recently, Ferreyra et al. \cite{ferreyra2024} have experimentally validated this model and discussed its shortcomings. The expressions for $\sigma_{zz}$ and $\sigma_{rr}$ at a given depth $h$ in the moving column are

\begin{align}
 \sigma_{zz}(h) =  &\frac{g \rhob \Ds}{4 B} [1 - {\rm e}^{-4 B h/\Ds}] \notag\\
 &+ \frac{4 g \Mp}{\pi \Ds^2} {\rm e}^{-4 B h/\Ds},  \label{eq-walterszz}
\end{align}
\begin{align}
 \sigma_{rr}(h) =  \frac{B}{\tan\phi} \sigma_{zz}(h), \label{eq-waltersrr}
\end{align}
where $\tan\phi$ is the effective friction corresponding to the wall yield locus \cite{NeddermanBook}, and $B/\tan\phi$ is the equivalent to the Janssen's force redirection factor $k$, which for discharging silos is
\begin{align}\label{eq-B}
 B &= \frac{\tan\phi \cos^2\delta}{(1+\sin^2\delta)-2y\sin\delta}, \\
 y &= \frac{2}{3c}[1-(1-c)^{3/2}],\notag\\
 c &= \frac{\tan^2\phi}{\tan^2\delta}.\notag 
\end{align}

Here, $\delta$ is the effective friction angle for the internal yield locus. As discussed by Nedderman (see \cite{NeddermanBook} \S 3.7), the wall yield locus $\phi$ should not be set simply as ${\rm atan}(\mu_{\rm wall})$. Instead, the Jenike's rule should be applied
\begin{equation}
 \tan\phi = \left\lbrace {\begin{array}{l} \mu_{\rm wall} \text{ if } \sin\delta \geq \mu_{\rm wall} \\ \sin\delta \text{ if } \sin\delta < \mu_{\rm wall} \end{array}}  \right. . \label{ec-jenike}
\end{equation}

\begin{figure}
 \begin{center}
  \includegraphics[width=0.9\columnwidth]{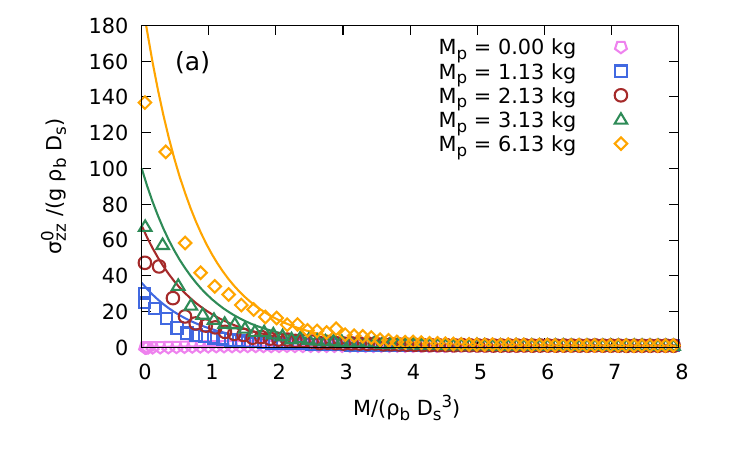}
  \includegraphics[width=0.9\columnwidth]{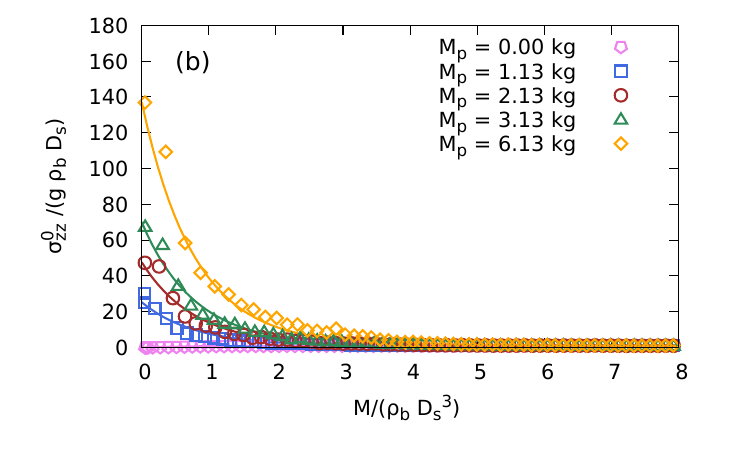}
 \end{center}
\caption{Normal pressure at he bottom of a discharging silo for different forcing overweight $\Mp$. Data points correspond to DEM simulations and solid lines correspond to: (a) the Walters model Eq.~(\ref{eq-walterszz}) and (b) corrected formula Eq.~(\ref{eq-walters2}) to account for the effect of the orifice by setting $\beta=0.7$. Simulations correspond to $\Ds=30d$, $\Do=6d$ and $\rho=2500$ kg/m$^3$.}
\label{fig:szz}
\end{figure}

Figure \ref{fig:szz}(a) shows the Walters model along with our simulation data. The particle--wall friction was set $\mu_{\rm wall}=0.5$ (as in  the simulations). The curves have been fitted by setting $\tan\delta=0.204$ and $\tan\phi=\sin\delta$, following Eq. (\ref{ec-jenike}). This corresponds to $\delta=11.5 ^\circ$ and $\phi= 11.3^\circ$ and leads to $B=0.25$. As we can see, the agreement is not perfect. However, as discussed in \cite{ferreyra2024}, we have to bear in mind that the Walters model does not take into account the presence of the orifice, which diminishes the pressure locally, affecting the mean pressure on the base. To account for this effect, we add a correction factor $\beta$ as

\begin{align}
 \sigma_{zz}(h) = & \frac{g \rhob \Ds}{4 B} [1 - {\rm e}^{-4 B h/\Ds}] \notag\\
 &+ \frac{\beta 4 g \Mp}{\pi \Ds^2} {\rm e}^{-4 B h/\Ds}. \label{eq-walters2}
\end{align}

By setting $\beta=0.7$ we obtain a better fit to the data (see Fig. \ref{fig:szz}(b)).

Since in Eq. (\ref{diff}) we also require the average $\langle \sigma_{rr} \rangle$ over the entire column of grains, we average the expression in Eq. (\ref{eq-walters2}) from $h=0$ to the total depth $h=z$ of the column, and use Eq. (\ref{eq-waltersrr}) to obtain

\begin{align}\label{eq-walters3}
\langle\sigma_{rr}\rangle (z) &= \frac{g \rhob \Ds}{4 \tan\phi} \\
&+ \frac{\Ds \left[1- {\rm e}^{-4 B z/\Ds} \right]}{4 z \tan\phi} \left( \frac{\beta 4 g M_{\rm p}}{\pi \Ds^2} - \frac{g \rhob \Ds}{4 B}\right). \notag 
\end{align}

\begin{figure}
 \begin{center}
  \includegraphics[width=0.9\columnwidth]{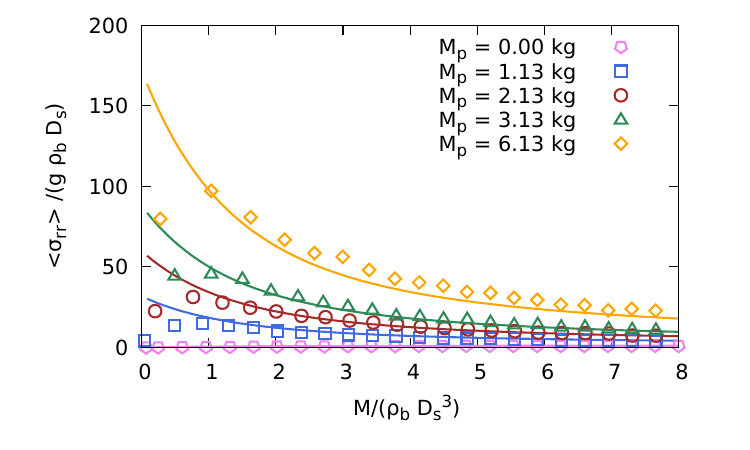}
 \end{center}
\caption{Mean normal pressure against the side walls as a function of the mass of grains in the silo during discharge with different forcing overweight. Symbols correspond to DEM simulations and solid lines to the modified Walters model Eq.~(\ref{eq-walters3}) with $\beta=0.7$. Simulations correspond to $\Ds=30d$, $\Do=6d$ and $\rho=2500$ kg/m$^3$.}
\label{fig:srr}
\end{figure}

In Fig. \ref{fig:srr} we show $\langle \sigma_{rr}\rangle$  obtained from DEM simulations along with the theoretical predictions from Eq.~(\ref{eq-walters3}). As we can see, the agreement is very good with the exception of the last part of the discharge, when $M(t) \lesssim 1.5 \rhob \Ds^3$ . When the height of the material remaining in the silo falls below $\approx \Ds$ the radial pressure decays to zero. This effect is not captured by the Walters model, not even after including the correction factor $\beta$. 

\subsection{Asymptotic pressure}

In section \ref{sec:results}, we have focused on the initial and final flow rates during discharge. These limiting values can be predicted using Eq. (\ref{diff}) by replacing the asymptotic limit of Eqs. (\ref{eq-walters2}) and (\ref{eq-walters3}). At the beginning of the discharge, for high columns (i.e., $z\gg\Ds$), these asymptotic expressions are

\begin{align}  \label{eq-asint-inf}
\sigma^0_{zz}(z\gg\Ds)&=\frac{g \rhob \Ds}{4 B}, \\
\langle\sigma_{rr}\rangle(z\gg\Ds)&=\frac{g \Ds}{4\tan\phi} \left[\rhob + \frac{1}{z} \left( \frac{\beta 4 M_{\rm p}}{\pi \Ds^2} - \frac{\rhob \Ds}{4B} \right) \right].\notag
\end{align}
At the end of the discharge, when $z\rightarrow 0 $, we have

\begin{align}  \label{eq-asint-0}
\sigma^0_{zz}(z\rightarrow 0)&=\frac{\beta 4 g M_{\rm p}}{\pi \Ds^2}, \\
\langle\sigma_{rr}\rangle(z\rightarrow 0)&=\frac{B}{\tan\phi} \frac{\beta 4 g M_{\rm p}}{\pi \Ds^2}. \notag 
\end{align}
We note that the asymptotic value for $\langle\sigma_{rr}\rangle$ does not agree with the simulation results as discussed in Fig.~\ref{fig:srr}. However, we will use this theoretical estimate due to the lack of alternative models in the literature.

\section{Initial flow rate: the Beverloo scaling}

In this section, we consider the flow rate $\Qi$ when the column height is very high ($z\gg \Ds$) at the beginning of discharge. By replacing the asymptotic Eq. (\ref{eq-asint-inf}) in Eq. (\ref{diff}), using $z = M(t)/(\rhob\As)$ and rearranging terms, we obtain

\begin{align} \label{eq-q-asint-inf}
 \Qi &= -\dot{M}(t) = \notag\\
 &= \frac{\pi \sqrt{2}}{4}\rhoo \sqrt{g} D_0^2 \left[ \frac{M(t)}{\rhob\As}\left(1-\frac{\muI}{\tan\phi}\right) \right. \notag\\
 &+ \frac{\Mp}{\rhob\As}\left(1-\frac{\muI\beta}{\tan\phi}-\muI\gamma\right) \\
 &+ \left. \frac{\muI \Ds}{B}\left(\frac{1}{4\tan\phi}-\frac{\alpha}{2}\right) + \frac{\muI \alpha \Do}{2 B}  \right]^{1/2}. \notag  
\end{align}

Equation (\ref{eq-q-asint-inf}) reduces to the Hagen--Beverloo expression for the flow rate if we select $\muI = \tan\phi = 1/(2\alpha) $, $B=1/4$ and $\gamma = (1-\beta)/\muI$. Under these conditions,

\begin{equation} \label{eq-q-asint2}
 \Qi = \frac{\pi \sqrt{2}}{4}\rhoo \sqrt{g} D_0^{5/2}.
\end{equation}

As we showed in the previous section, the Walters expressions for the pressure fit the DEM data if $\tan\delta=0.204$. Then $\tan\phi=0.20$. Equation (\ref{eq-B}) then yields $B=1/4$, which automatically fulfills one of the necessary conditions.  Therefore, the well known Hagen's relation will hold if we take $\muI=0.20$ and $\alpha=2.5$. Since we set $\beta=0.7$ to fit the DEM pressure due to the presence of the orifice, the above relations require $\gamma =1.5$. We note that these values are independent of particular silo conditions, such as the specific values for $\Ds$, $\Do$, and the particle--particle or particle--wall interaction. The result of Eq.~(\ref{eq-q-asint2}) is equivalent to the one previously found in the case of a free discharge \cite{darias2020}.

When calculating our energy contributions, we assumed that the system can be considered as a continuum and the size of the grains $d$ is much smaller than any other length scale in the system. This is one of the assumptions of the local $\mu(I)$-rheology. This approximation is not entirely reasonable when $\Do$ is only a few times $d$. Although the effect of particle size can be introduced by means of a non-local rheology approach \cite{dunatunga2022}. Here, we follow the simplistic correction, in the style of Beverloo, made by replacing the orifice diameter $D_{\rm o}$ by an effective smaller diameter $D_{\rm o}-kd$. In our simulations, we can fit the data for free discharge using Eq.~(\ref{eq-q-asint2}) with $k = 1.72$. We will use this value to correct $\Do$ in the rest of the model equations in this work. 

As discussed in \cite{darias2020}, we can replace $\rho_{\rm o}\approx \rhob/2$ in Eq. (\ref{diff}). This is based on measurements carried out on our DEM simulations. Then, in Eq. (\ref{diff}), the non-dimensional prefector $\frac{\pi\sqrt{2}}{8} \approx 0.56$ can be compared with the constant $C$ in the Beverloo rule. This value deviates less than $4 \%$ from the universal value $C=0.58$ obtained by fitting the experimental data \cite{NeddermanBook}.

In Figs.~\ref{fig:presion} to \ref{fig:silo} we have included expression (\ref{eq-q-asint2}) --- with the replacements described in the previous paragraph for $\Do$ and $\rhoo$--- for comparison with the data for the initial flow rate. The agreement is excellent. This is in line with a similar development presented in \cite{darias2020} for freely discharging silos. 

\section{Final flow rate: {inviscid} fluid scaling}
\label{sec:final-flow}

We now focus on the flow rate $\Qf$ at the end of the discharge, when the column of grains vanishes. Replacing the asymptotic Eq. (\ref{eq-asint-0}) in Eq. (\ref{diff}) we obtain
 \begin{align} \label{eq-q-inf}
\Qf &= -\dot{M}(t) = \notag\\
    &= \frac{\pi \sqrt{2}}{4} \rhoo \sqrt{\frac{g \Mp}{\rhob \As}} \Do^2 \\
    &\times \left[ 1 - 2\muI\alpha\beta + 2\muI\alpha\beta \frac{\Do}{\Ds} - \muI\gamma \right]^{1/2}. \notag
\end{align}
If, as before,  we set $\rhoo\approx\rhob/2$ and use the relations and values obtained in the previous section for $\muI$, $\alpha$, $\beta$ and $\gamma$, then
\begin{equation} \label{eq-q-0}
 \Qf = \frac{\pi}{8} \sqrt{2 \rhob P_{\rm p}} \Do^{2} \left(\beta\frac{\Do}{\Ds}\right)^{1/2},
\end{equation}
where we defined the pressure applied by the piston as $P_{\rm p} = g\Mp/\As$. This equation resembles that of an inviscid fluid encountering a plate with an orifice of diameter $\Do$ in a pipe of diameter $\Ds$ \cite{Reader2015}. 

Equation (\ref{eq-q-0}) predicts that the flow rate depends on the applied piston pressure as a regular fluid. This scaling agrees with the simulation data for high applied pressures (see blue solid line in Fig.~\ref{fig:presion}). However, it is clear that for zero applied pressure, the flow rate cannot drop to zero as suggested by Eq.~(\ref{eq-q-0}) but must tend to the flow rate of a free discharge. The dependency of $\Qf$ on the density predicted by Eq.~(\ref{eq-q-0}) is also consistent with the DEM data, as shown in  Fig.~\ref{fig:densidad}. 

\begin{figure}
    \centering
    \includegraphics[width=0.5\columnwidth]{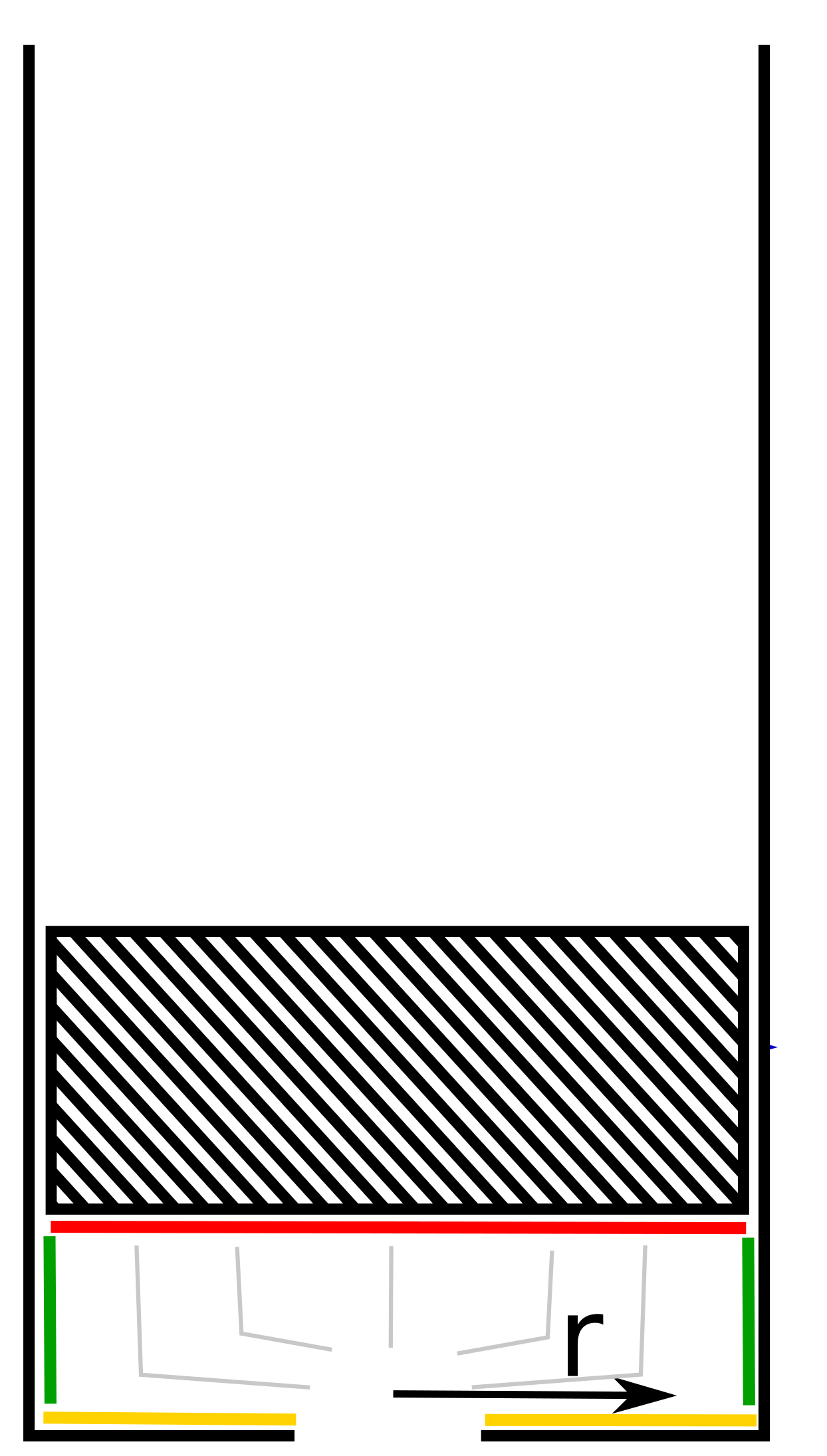}
    \caption{Streamlines when the overweight approaches the base.}
    \label{fig:lineas2}
\end{figure}

Despite the success of  Eq.~(\ref{eq-q-0}) in predicting the pressure and density effect, which contrasts with those observed in free discharges, the dependency on $\Do$ and $\Ds$ is not confirmed by the simulation data. While Figs.~\ref{fig:orificio} and \ref{fig:silo} show that $\Qf$ scales as $\Do^3$ and $\Ds^{-1}$, Eq.~(\ref{eq-q-0}) predicts  $\Do^{5/2}$ and $\Ds^{-1/2}$. The origin of this failure in the model can be related to the actual flow pattern at the end of the discharge. The conical flow pattern depicted in Fig.~\ref{fig:lineas1} can be significantly altered when the piston approaches the base (see Fig.~\ref{fig:lineas2}). Also, as we observed in Fig.~\ref{fig:srr}, the available model for the radial stress fails to predict the behavior at the end of the discharge. Moreover, the approximation that $W_{\rm el}=0$ can be an important missing contribution in the theory, specially for the final stages of the discharge. Developing a fully successful model for the final stages of a forced discharge remains an open challenge.

\section{Conclusions}

We have presented DEM simulations of forced silo discharges where we vary material density, silo diameter, outlet diameter, and forcing pressure. The flow rate at the beginning of the discharge for a tall column of grains agrees with that of a free discharge. However, at the end of the discharge, the flow rate accelerates and the scaling with the main control variables differs from that of a free discharge. In particular, the final flow rate scales with applied pressure and material density as in a viscous fluid. In addition, the silo diameter becomes a relevant parameter since the flow rate is inverse to it. More surprising yet is the cubic scaling observed with the outlet diameter.  

A model based on energy balance and local $\mu(I)$-rheology allowed us to predict the flow rate at the beginning of the discharge. However, this model shows only partial qualitative agreement for the final flow rate. Although the pressure and density scaling are well captured by the model, the effect of silo diameter and discharge outlet diameter is poorly predicted. One may speculate that more sophisticated rheology models may be necessary. However, the usual extension to non-local rheology does not seem to be the natural pathway in these highly forced systems where rapid flow is ubiquitous. These strongly forced granular flows pose new challenges to our understanding of the response of granular materials.

\begin{acknowledgements}
We are grateful for discussions with María Victoria Ferreyra and Diego Maza. This work has been supported by CONICET (Argentina) through grant PIP-717, UNLPam (Argentina) through grant F64, ANPCyT (Argentina) through grant PICT-2020-SERIE-A–02611 and UTN (Argentina) through grant PID-9875. 
\end{acknowledgements}

\section*{Appendix A}

Following Walters \cite{walters1973}, the normal $\sigma_{rr}$ and tangential $\sigma_{rz}$ stresses in a silo can be put in terms of the vertical stress $\sigma_{zz}$ according to

\begin{align}
    \sigma_{rr}&= \frac{B}{\tan \phi} \sigma_{zz}, \notag\\
    \sigma_{rz}&= B \sigma_{zz}.
\end{align}
Here, $\phi$ is the wall friction angle and $B$ is a constant given in Eq.~(\ref{eq-B}) that depends on $\phi$ and the internal angle of friction $\delta$ (see section \ref{sec:walters} for details).

Since all components of the stress tensor are proportional to $\sigma_{zz}$, the shear stress along any plane will be proportional to $\sigma_{zz}$. For the plane of the stagnant cone (see yellow line in Fig.~\ref{fig:lineas1}) that forms an angle $\omega$ with the silo axis we obtain

\begin{align}\label{eq:shear-stress-cone}
    \sigma_{r\theta}&=\muI \sigma_{\theta\theta} \\
    &= \muI \sigma_{zz} \notag\\
    &\ \ \ \ \times\left[ \frac{B}{\tan\phi}\cos^2\omega - 2B \sin\omega\cos\omega+\sin^2\omega \right],\notag 
\end{align}
where we have used the $\mu(I)$-rheology constitutive relation with $\sigma_{\theta\theta}$ that normal stress on the plane of the stagnant cone in spherical coordinates. 

Equation (\ref{eq:shear-stress-cone}) is valid next to the silo vertical walls. Hence, $\sigma_{zz}$ should be taken at the height where the stagnant cone meet the side wall. However, as a proxy, we can simply take the vertical stress $\szzo$ at the bottom of the silo and let the factor $\alpha$ defined in the main text to compensate for the deviation.


\begin{thebibliography}{50}
\bibitem{foot} Note that we have selected the negative root since this is the physically meaningful root corresponding to negative $\dot{M}$.
\end{thebibliography}
\end{document}